\title{A Criterion for Attaining the Welch Bounds with Applications for Mutually Unbiased Bases}   
\author{Aleksandrs Belovs$^1$\thanks{Research supported in part by MITACS and Province of Ontario.} and Juris Smotrovs$^2$\thanks{Supported by the European Social Fund and by the University of Latvia research project No Y2-ZP14-100. 
}}

\documentclass{article}[10pt]   
\usepackage{amsmath, amssymb}   
\usepackage{graphicx}

\newtheorem{thm}{Theorem}   
\newtheorem{lem}[thm]{Lemma}   
\newtheorem{col}[thm]{Corollary}   
\newtheorem{prop}[thm]{Proposition}   
\newtheorem{open}[thm]{Open Problem}   
   
\newcommand{\pfstart}{{\em Proof. \quad}}   
\newcommand{\pfend}{$\blacksquare$\par\addvspace{12pt plus1pt minus3pt}}   

\newcommand{\rank}{\mathop{\rm rank}\nolimits}   
\newcommand{\tr}{\mathop{\rm Tr}\nolimits} 
\newcommand{\diag}{\mathop{\rm diag}\nolimits} 

\begin{document}   
\maketitle   
\begin{center}   
{\small $^1$ Department of Combinatorics and Optimization, and Institute for Quantum Computing,\\ University of Waterloo, Waterloo, Ontario, Canada, N2L 3G1}\\
{\small $^2$ Department of Computer Science, University of Latvia, Rai\c{n}a bulv\={a}ris 19,  R\={\i}ga, Latvia}\\
\vspace{0.3cm}
{\small stiboh@inbox.lv, juris.smotrovs@lu.lv}\\   
\vspace{0.8cm}   
\end{center}   

\begin{abstract}   
The paper gives a short introduction to mutually unbiased bases and the Welch bounds and demonstrates that the latter is a good technical tool to explore the former. In particular, a criterion for a system of vectors to satisfy the Welch bounds with equality is given and applied for the case of MUBs. This yields a necessary and sufficient condition on a set of orthonormal bases to form a complete system of MUBs. 

This condition takes an especially elegant form in the case of homogeneous systems of MUBs. We express some known constructions of MUBs in this form. Also it is shown how recently obtained results binding MUBs and some combinatorial structures (such as perfect nonlinear functions and relative difference sets) naturally follow from this criterion.

Some directions for proving non-existence results are sketched as well.
\end{abstract}   

\section{Mutually Unbiased Bases}   
The current research originated in the problem of constructing a complete set of mutually unbiased bases and is inspired mostly by~\cite{are}.    

A set of {\em mutually unbiased bases} (MUBs) in the Hilbert space $\mathbb{C}^n$ is defined as a set of orthonormal bases $\{B_0, B_1,\dots, B_r\}$ of the space such that the absolute value of a scalar product $|\langle x|y\rangle|$ is equal to $\frac{1}{\sqrt{n}}$ for any two vectors $x\in B_i$, $y\in B_j$ with $i\ne j$. For the sake of brevity we will further call the absolute value of a scalar product of two vectors as the {\em angle} between these vectors. We will often group vectors of a basis into a matrix and say that two unitary matrices are mutually unbiased iff the bases obtained from their columns are. Bases with such properties were first observed by Schwinger in~\cite{schwinger}. The name of mutually unbiased bases is due to Fields and Wootters~\cite{wootters}. 

The applications of MUBs include quantum state determination~\cite{ivanovic, wootters}, quantum cryptography (the protocol BB84 due to Bennet and Brassard~\cite{bennet} is a classical example of such a usage), the Mean King's problem~\cite{aharonov} and Wigner functions~\cite{wigner}. A good source of an up-to-date information on MUBs can be found on~\cite{problem13}.   

Clearly, if $n=1$ then one unit vector (in fact scalar) repeated necessary amount of times gives a set of MUBs of any size. This result does not seem extremely useful, so we will further assume the dimension of the space $n$ is at least 2. In this case it can be proved that the number of bases in any set of MUBs in $\mathbb{C}^n$ doesn't exceed $n+1$ (see Theorem~\ref{prevywenie} later in the text). A set of bases that achieves this bound is called a {\em complete set of MUBs}. An interesting question is whether such a set exists for any given dimension $n$. The answer is positive if $n$ is a prime power~\cite{ivanovic, wootters}. The corresponding constructions are listed in section~\ref{known} of this paper. In all other cases (even for $n=6$) the question is still open, despite a considerable effort spent on solving this problem (see, e.g.,~\cite{six}).   

The search for complete systems of MUBs is complicated because of the number of bases we should find and because of the non-obviousness of the value of the angle $\frac1{\sqrt{n}}$. Using the Welch bounds (described in the next section) we give a sufficient and necessary condition that uses solely orthogonality of vectors. Clearly, it is a much more studied and intuitive relation. 

This is not the first attempt to substitute the angle $\frac1{\sqrt{n}}$ by zero. An alternative approach appears in the classical paper~\cite{wootters}: 
\begin{prop} 
\label{density} 
Consider the operation that maps a state $|x\rangle\in\mathbb{C}^n$ to the corresponding traceless density matrix $Y_x = |x\rangle\langle x| - I/n$. Then $|\langle x|y\rangle|=\frac1{\sqrt{n}}$ if and only if matrices $Y_x$ and $Y_y$ are orthogonal with respect to the trace inner product: $Tr(Y_x^\dagger Y_y)=0$.  
\end{prop} 

In particular, applications of MUBs in quantum state tomography are based on this observation. 

Our approach is slightly different. From the collection of $n+1$ orthonormal bases in $\mathbb{C}^n$, pretending to be mutually unbiased, we extract $n$ flat (with all entries having the same absolute value) vectors, each in $\mathbb{C}^{n^2}$. Next, from each pair of these vectors we obtain a new vector from the same space. We prove that the bases of the original collection are MUBs if and only if the latter vectors are pairwise orthogonal. It is not a problem to find ${n\choose 2}$ orthogonal flat vectors in $\mathbb{C}^{n^2}$, but, in general, they won't be decomposable back to pairs. 

Moreover, if we restrict our attention to homogeneous systems of MUBs (see section~\ref{odnorod} for the definition), it is possible to reduce the criterion to only two matrices from $\mathbb{C}^n$ and orthogonality conditions obtained in a similar fashion. In order to show the usability of our result we show how it sheds light on the known constructions of complete sets of MUBs. In particular, we give a bit easier proofs that these constructions do result in complete sets of MUBs. 

We also show how this approach naturally leads to some applications of combinatorial structures to MUBs that were obtained recently. In particular, we extend the correspondence between planar functions and splitting semiregular relative difference sets to the case of non-splitting ones.

\section{Welch bounds and Crosscorrelation}   
Welch bounds are the inequalities from the following theorem: 
\begin{thm}   
\label{neravenstvo}   
For any finite sequence $\{x_i\}$ of vectors in Hilbert space $\mathbb{C}^n$ and any integer $k\ge 1$ the following    
inequality holds: 
\begin{equation}   
\label{welch}   
{n+k-1\choose k} \sum_{i,j} |\langle x_i | x_j \rangle|^{2k} \ge  \left(\sum_i \langle x_i|x_i\rangle^k\right)^2. 
\end{equation} 
\end{thm}  

The proof will be given in section~\ref{criterion}, but for now let us note that these inequalities were first derived (in the case of all vectors having the same norm) by Welch in~\cite{welch}. It is worth to become acquainted with his motivation. 

In order to do this we should define sequences with low correlation. For a systematic treatment of the topic see~\cite{correlation}. Let $u$ and $v$ be complex periodic sequences of equal period $n$. Usually the sequences are defined as $u_i = \omega_n^{a_i}$ with $a_i$ from $\mathbb{Z}_n$. ($\omega_n$ is a primitive $n$-th power root of unity: $\omega_n = e^{\frac{2\pi i}n}$, $\mathbb{Z}_n$ is the ring of integers modulo $n$). The binary case (with $n=2$) is the most common. The {\em (periodic) correlation} of $u$ and $v$ is defined as (where $L$ stands for the left cyclic shift function) 
$$\theta_{u,v}(\tau) = \langle L^\tau(u)|v\rangle = \sum_{i=1}^n \overline{u_i}v_{i+\tau}.$$ 
The correlation of a sequence with itself is called its {\em autocorrelation} $\theta_u(\tau) = \langle L^\tau(u)|u\rangle$. The correlation of two shift-distinct sequences is usually called {\em crosscorrelation}. 

Informally, the correlation of binary sequences characterizes the number of places two sequences coincide minus the number of places they differ. For random sequences magnitude of this value is small, so it can be used as a measure of the pseudorandomness of a sequence. The correlation is called {\em ideal} if it is as small as possible (0 or $\pm1$). It is considered low, if it is $O(\sqrt{n})$ (an expected value for random sequences). For example, {\em m-sequences} (the maximal length sequences generated by a linear feedback shift register (LFSR)) have ideal autocorrelation, since for them $\theta(\tau)=-1$ for any $\tau\not\equiv0\pmod{n}$. This, among other properties, explains why they are used in cryptography (as a main building block of nearly every stream cipher) and electronic engineering (e.g., in radars). 

Families of sequences with low crosscorrelation are also well-studied. A nice property of these sequences is that they can be transmitted through the same channel simultaneously without mutual disturbance. By the time Welch was writing his paper there were some good families of sequences with low auto- and cross-correlation and he got interested in obtaining upper bounds on the number of sequences in a family. 

For example, one classical family of sequences was proposed by Gold in~\cite{gold}. For any integer $n$ he constructed a family of $2^n+1$ binary sequences of period $2^n-1$ and correlation between any two of them takes only three possible values: $-1, -(2^{(n+1)/2}+1)$ and $2^{(n+1)/2}-1$. 

Similarity of this family and a complete family of MUBs is apparent. Both are built of vectors from $\mathbb{C}^n$, vectors are joined in blocks of size $n$, the number of blocks is approximately the same and the ratios of possible inner products and norms of vectors also almost agree. So, an attempt to apply Welch bounds to the problem of MUBs seems quite reasonable. 

Even more, it turns out that Alltop in his work~\cite{alltop} of 1980 (i.e., one year before the work~\cite{ivanovic} of Ivanovi\'c) for any prime $p\ge5$ gave a set of $p$ sequences with period $p$ and elements with absolute value $\frac1{\sqrt{p}}$, such that the crosscorrelation is given by 
$$|\theta_{uv}(\tau)| = \left\{\begin{array}{rcl} 
1&,&u=v\mbox{ and } \tau=0;\\ 
0&,&u\ne v\mbox{ and } \tau=0;\\ 
\frac1{\sqrt{p}}&,& \tau\ne0. \end{array} \right. $$ 
Clearly, these sequences with different shifts and the standard basis give a complete set of MUBs in $\mathbb{C}^p$. This result was generalized to prime power dimensions in~\cite{constructions}.

\section{Link between MUBs and the Welch Bounds} 
In our first application of the Welch bounds to MUBs we can apply the original approach of Welch in the new settings. It is easy to check that a union of orthonormal bases satisfy the Welch bound for $k=1$ (it can be done either directly using~(\ref{welch}) or using Theorem~\ref{kriterij} further in the text). So, we should use $k=2$.  

\begin{thm} 
\label{prevywenie} 
If $n\ge 2$ then the maximal number of mutually unbiased bases in ${\mathbb C}^n$ does not exceed $n+1$.  
\end{thm} 

\pfstart 
Suppose we have a system of $n+2$ MUBs. Join all vectors of the system into one big sequence $\{x_i\}$ of size $n(n+2)$. Let us fix $k=2$ and calculate the left hand side of~(\ref{welch}). We have $n(n+2)$ vectors, each giving the scalar product 1 with itself and $n(n+1)$ scalar products of absolute value $\frac1{\sqrt{n}}$ with vectors from other bases. Summing up, we have: 
$${n+1 \choose 2}\sum_{i,j} |\langle x_i | x_j \rangle|^4 = \frac{n(n+1)}2 \left[n(n+2)\left(1 + n(n+1)\cdot\frac1{n^2}\right)\right] = \frac{n(n+1)(n+2)(2n+1)}2.$$   
For the right hand side we have: 
$$\left(\sum_i \langle x_i|x_i\rangle^2\right)^2 = n^2(n+2)^2 > \frac{n(n+1)(n+2)(2n+1)}2,$$ 
in a contradiction with the Welch bound for $k=2$.  
\pfend 

Originally it was proved in~\cite{wootters} using the result of Proposition~\ref{density}. 

If we reduce the number of MUBs from $n+2$ to $n+1$ we don't get an apparent contradiction. However, even in this case the Welch bounds prove themselves to be useful. Klappenecker and R\"otteler seem to be the first ones to realize this by proving the `only if' part of the following theorem in~\cite{are}. The `if' part seems first to appear later, in ~\cite{roy}. 

\begin{thm} 
\label{MUB2design}   
Let $\{B_i\}$ be a set of $n+1$ orthonormal bases in an $n$-dimensional Hilbert space and $X$ be the union of these bases (that is the sequence of vectors, each of them appearing in the sequence the same number of times it appears in the bases). Then $X$ satisfies the Welch bound for $k=2$ with equality if and only if $\{B_i\}$ form a complete system of MUBs.   
\end{thm}  

\pfstart  
If $\{B_i\}$ is a complete system of MUBs and $X=\{x_i\}$ is the union of its bases, then calculations similar to ones in the proof of Theorem~\ref{prevywenie} show  
$${n+1\choose 2}\sum_{i,j} |\langle x_i | x_j \rangle|^4 = \frac{n(n+1)}2\left[n(n+1)\left(1 + n^2 \cdot\frac1{n^2}\right)\right] = n^2(n+1)^2$$   
and   
$$\left(\sum_i \langle x_i|x_i\rangle^2\right)^2 = n^2(n+1)^2.$$   

And vice versa, suppose $X$, being a union of orthonormal bases, attains the Welch bound for $k=2$. Then, $|\langle x|x\rangle|^4=1$ for each $x$ in $X$, $|\langle x|y\rangle|^4 = 0$ for two different vectors of the same basis, and by the inequality between square and arithmetic means we get:   
$$\sum_{x\in B_i} |\langle x | y\rangle|^4 \ge \frac1n \left(\sum_{x\in B_i} |\langle x|y\rangle|^2 \right)^2 = \frac1n.$$   
for any vector $y$ of unit length. To attain the Welch bound, this inequality must actually be an equality, which is achieved only if $|\langle x|y\rangle|^2$ has the same value for all vectors $x$ from $B_i$. This means that bases $\{B_i\}$ form a complete system of MUBs.   
\pfend   

Systems of vectors attaining the Welch bounds have been investigated before. A system of complex vectors from $\mathbb{C}^n$ attaining the Welch bounds for all $k\le t$ is called a {\em complex projective $t$-design}. This is a Chebyshev-type averaging set on the n-dimensional complex unit sphere $\mathbb{C}S^{n-1}$, in the sense that the integral of every polynomial of degree $\le t$ is equal to the average of its values on the vectors from the $t$-design. See~\cite{are} for more details. 

We give a criterion for attaining the Welch bounds in the next section. 

\section{Criterion for Attaining the Welch Bounds} 
\label{criterion} 
Let us at first define the Schur product of two matrices. Let $A=(a_{ij})$ and $B=(b_{ij})$ be two matrices of equal sizes. The {\em Schur product} (see, for example, chapter 7 of~\cite{horn}) is the matrix of the same size (denoted by $A\circ B$) with its $(i,j)$-entry equal to $a_{ij}b_{ij}$. In other words, multiplication is performed component-wise. The $k$-th Schur power of the matrix $A$ is again the matrix of the same size (denoted by $A^{(k)}$) with its $(i,j)$-entry equal to $a_{ij}^k$. 

Additionally, we shall use notation $A^\dagger$ for the adjoint matrix (complex conjugated and transposed) and the term {\em self-adjoint} for matrices $A$ satisfying $A^\dagger = A$ (also called Hermitian).  

We will at first give a proof of the Welch bounds and then extract the equality criterion from the proof.  

\vspace{0.2cm} 
{\em Proof of Theorem~\ref{neravenstvo}. \quad} 
Let us construct the Gram matrix $G=(a_{ij})$ with $a_{ij}=\langle x_i | x_j\rangle$ and consider its $k$-th Schur power $G^{(k)}$ (with $k$ being a  positive integer). The square of its Euclidean norm (see chapter 5 of~\cite{horn}, also known as Frobenius norm and Schur norm) is defined by   
\begin{equation}   
\label{Adef}   
\left(\|G^{(k)}\|_E\right)^2 = \sum_{i,j} |\langle x_i | x_j \rangle|^{2k} =    
\end{equation}   
Unitary operators, applied both from the left and the right, do not change the Euclidean norm of a matrix. Any self-adjoint matrix can be transformed into a diagonal matrix with real entries (its eigenvalues) on the diagonal by a unitary transformation, and $G^{(k)}$ is a self-adjoint matrix, hence   
\begin{equation}   
\label{Nerav1}   
= \sum_{\lambda\in \sigma(G^{(k)})} \lambda^2 \ge   
\end{equation}   
here $\sigma$ is the spectrum (the multiset of the eigenvalues of a matrix). By the inequality between square and arithmetic means, we have (let us remind that the rank of a self-adjoint matrix is equal to the number of its non-zero eigenvalues):   
\begin{equation}   
\label{Nerav2}   
\ge \frac{1}{\rank(G^{(k)})}(\tr G^{(k)})^2 \ge \frac{1}{{n+k-1\choose k}} \left(\sum_i \langle x_i|x_i\rangle^k\right)^2.   
\end{equation}   
The last estimation on the rank of $G^{(k)}$ we will prove later.   
\pfend   

\begin{thm}   
\label{kriterij}
Let $B$ be a matrix and $X\subset\mathbb{C}^n$ be the sequence of its columns. Let $w_1,w_2,\dots,w_n$ be the rows of the matrix. Then $X$ attains the Welch bound for a fixed $k$ if and only if all vectors from   
$$W = \left\{\sqrt{{k \choose k_1,\dots,k_n}}w_1^{(k_1)}\circ w_2^{(k_2)}\circ\cdots\circ w_n^{(k_n)} \mid k_i\in\mathbb{N}_0, k_1+k_2+\cdots+k_n=k\right\}$$   
are of equal length and pairwise orthogonal.   
\end{thm}   

In other words, each vector of $W$ is a Schur product of a $k$-multiset of rows of $B$ with a coefficient that is the square root of the multinomial coefficient of the multiset    
$${k \choose k_1,\dots,k_n} = \frac{k!}{k_1!k_2!\cdots k_n!}.$$   

\pfstart   
At first, let us note that matrix $G$ in~(\ref{Adef}) is equal to $B^\dagger B$. So (if each $w_i$ is treated as a row vector):   
$$G = w_1^\dagger w_1 + w_2^\dagger w_2\cdots + w_n^\dagger w_n.$$   
By the formula for a power of a sum, we obtain
$$G^{(k)} = \sum_{k_1+k_2+\cdots+k_n=k} {k \choose k_1,\dots,k_n} \left(w_1^{(k_1)}\circ w_2^{(k_2)}\circ\cdots\circ w_n^{(k_n)}\right)^\dagger\left(w_1^{(k_1)}\circ w_2^{(k_2)}\circ\cdots\circ w_n^{(k_n)}\right).$$   
In other words, $G^{(k)} = C^\dagger C$, where the rows of $C$ are exactly the vectors from $W$. This gives the bound on the rank of $G^{(k)}$ used in formula~(\ref{Nerav2}), because the number of $k$-multisets of an $n$-set equals ${n+k-1\choose k}$ (see, e.g., section 1.2 of~\cite{stanley}).   

By observing the inequality between~(\ref{Nerav1}) and~(\ref{Nerav2}), we see that $X$ satisfies the Welch bound for a fixed $k$ with equality if and only if $G^{(k)}$ has ${n+k-1\choose k}$ equal non-zero eigenvalues (all other eigenvalues are automatically zeros due to the rank observations).   

It is a well-known fact that for any matrices $P$ and $Q$ the set of non-zero eigenvalues of matrices $PQ$ and $QP$ are equal whenever these two products are defined (see section 1.3 of~\cite{horn}). Hence, $CC^\dagger$ have ${n+k-1\choose k}$ equal non-zero eigenvalues, and because it is a self-adjoint matrix of the same size it is a scalar multiple of the identity matrix. And the latter is equivalent to the requirement on the set $W$.   
\pfend   

We haven't hitherto seen the pair of theorems~\ref{neravenstvo} and~\ref{kriterij} appearing in such a general form, however all ideas involved in the proof have already appeared in the proofs of other results. As we have already said, Welch was the first who derived the bounds~(\ref{welch}) in the case when all vectors have unit norm and $k$ is arbitrary. It was done in~\cite{welch}. The variant of theorem~\ref{kriterij}, with $k=1$ and all vectors of equal length, seems first to appear in~\cite{massey}. Our proof is a generalization of an elegant proof found in~\cite{waldron}. In the latter paper the Welch bounds are stated in the case of vectors of different length, but it deals with the case of $k=1$ only.   

\section{Application of the Criterion to MUBs} 

At first let us state the following easy consequence of theorem~\ref{kriterij}: 

\begin{col}   
\label{design2W}   
Let $B$ be a matrix and $X\subset\mathbb{C}^n$ be the sequence of its columns. Let $w_1,w_2,\dots,w_n$ be the rows of the matrix. Then $X$ satisfy the Welch bound for $k=2$ with equality if and only if all vectors from $W = \{w_i^{(2)}\}\cup\{\sqrt{2}w_i\circ w_j\mid\ 1\le i < j\le n\}$ are of equal length and pairwise orthogonal.   
\end{col}   

Suppose we have a complete system of MUBs: $\{B_0, B_1, \dots, B_n\}$. We can always represent them in the first basis $B_0$, thus we can assume that the first basis is the standard basis (the identity matrix). Then the matrices representing all other bases have all their entries equal by the absolute value to $\frac{1}{\sqrt{n}}$. 

A matrix with complex entries and with all entries having the same absolute value is called a {\em flat} matrix. If it is additionally unitary, it is called a {\em complex Hadamard matrix}. It is common to rescale flat matrices in such a way that each its element has absolute value 1. We will usually assume that. In the case of an $n\times n$ complex Hadamard matrix it is sometimes more convenient to assume each element having absolute value $\frac1{\sqrt{n}}$, sometimes 1. According to the situation we will use both assumptions, it will be usually clear from the context what is meant.  

Complex Hadamard matrix is a generalization of classical Hadamard matrix that satisfies the same requirements, but with all entries real (i.e., $\pm1$) (see, for example, section I.9 of~\cite{hadamard}). We will further use term Hadamard matrix or just Hadamard to denote complex Hadamard matrices. 

Two Hadamard matrices are called {\em equivalent} if one can be got from the other using row and column multiplications by a scalar and its permutations. Some classes of equivalent Hadamards are classified. See~\cite{tadey} for more details.

A system of Hadamards such that any two are mutually unbiased is called a {\em system of mutually unbiased Hadamards} or {\em MUHs} for short. The following result is obvious 
\begin{prop} 
A complete system of MUBs exists in space $\mathbb{C}^n$ if and only if there is a system of $n$ MUHs in the same space. 
\end{prop} 
A system of $n$ MUHs in $\mathbb{C}^n$ is called a {\em complete system of MUHs}.  We will turn to the investigation of complete systems of MUHs in the remaining part of the paper.  

Now we are able to prove the following theorem:   
\begin{thm}   
\label{Glavnaja}   
Let $\{B_i\}$ ($i=1,2\dots,n$) be a set of $n$ Hadamards in $\mathbb{C}^n$ and $B$ be a concatenation of these matrices (i.e. a $n^2\times n$-matrix having as columns all columns appearing in $\{B_i\}$). Then $\{B_i\}$ form a complete set of MUHs if and only if all vectors from $W'=\{w_i\circ w_j\mid\ 1\le i \le j\le n\}$ are pairwise orthogonal, where $\{w_i\}$ are the rows of $B$.   
\end{thm}   
\pfstart   
Let us denote the $n\times n$ identity matrix by $B_0$. By Theorem~\ref{MUB2design}, we see that the set $\{B_0,B_1,\dots,B_n\}$ is a complete set of MUBs if and only if the set of columns of all these matrices attains the Welch bound for $k=2$. Now from Corollary~\ref{design2W} it follows that it only remains to show that vectors from $W$, as it was defined in Corollary~\ref{design2W}, are of equal length and orthogonal if vectors of $W'$ are orthogonal.   

If a vector from $W$ is multiplied by itself using the Schur product, the result has one 1 and all other entries equal to 0 in the part corresponding to $B_0$, and all entries in other parts by the absolute value are equal to $\frac{1}{n}$. Hence, the length of the vector is $\sqrt{1+n^2\frac{1}{n^2}} = \sqrt{2}$.    

If two distinct vectors are multiplied, the result has only zeroes in the first part, and its length is $\sqrt{n^2\frac{1}{n^2}}=1$. We see that all vectors from $W$ have the same length. 

Moreover, the part of $B_0$ contributes zero to the inner product of 2 distinct vectors of $W$, hence vectors of $W$ are orthogonal if and only if the corresponding vectors of $W'$ are.   
\pfend   

Let us restate the last theorem. Suppose $B$ is a flat $n\times n$-matrix. Construct the weighted graph $K(B)$ as follows. Its vertices are all multisets of size 2 from $\{1,...,n\}$. Semantically a vertex $\{i,j\}$ represents the Schur product of the $i$-th and the $j$-th row of $B$. The weight of an edge is the inner product of the vertices it joins. (Of course, thus defined, the weight depends on the order of the vertices, but let us fix a direction of each edge, say lexicographical). Then theorem~\ref{Glavnaja} can be restated by saying that Hadamards $B_1,\dots,B_n$ form a set of MUHs in $\mathbb{C}^n$ if and only if the sum of weights of each edge in all of  $K(B_1),\dots,K(B_n)$ equals 0. In fact, there is no need to consider edges between vertices that have an element in common, since they will be orthogonal. 

It does not seem that this restatement makes the problem much easier comparing to the initial formulation. However, careful examination of the possible configurations of weights that can be achieved in $K(B)$ may shed some light on the problem. In the next section we consider a special case of systems of MUHs for which Theorem~\ref{Glavnaja} yields a considerable simplification. 

\section{Homogeneous Systems of MUBs} 
\label{odnorod} 
Suppose we have a flat $n\times n$-matrix $A=(a_{i,j})$ and an Hadamard matrix $H=(h_{i,j})$ of the same dimensions. Consider the following system of Hadamards (it's assumed that each element of $A$ and $H$ has absolute value 1)
\begin{equation} 
\label{homogenous} 
(v^{(r)}_{k})_\ell = \frac{1}{\sqrt{n}} a_{\ell, r} h_{\ell, k} 
\end{equation} 
with $r$ being a matrix index, $k$ being a column index and $\ell$ being a row index ($r,k,\ell\in\{1,\dots,n\}$). In other words, the $i$-th matrix is given by $\diag(v_i)H$, where $v_i$ is the $i$-th column of $A$. We will call such a set of Hadamards (in the case it forms a set of MUHs) a {\em homogeneous system of MUHs}, or a {\em homogeneous system of MUBs} if the identity matrix is appended. The name is borrowed from~\cite{six}. 

From theorem~\ref{Glavnaja} it follows that the system from~(\ref{homogenous}) forms a system of MUHs if and only if 
$$\langle w_{\ell_1}\circ w_{\ell_2}| w_{\ell_3}\circ w_{\ell_4}\rangle = \frac1n \sum_{r,k} \overline{a_{\ell_1,r} h_{\ell_1,k} a_{\ell_2, r} h_{\ell_2,k}} a_{\ell_3,r} h_{ \ell_3,k} a_{\ell_4,r} h_{ \ell_4,k} =$$ 
$$= \frac1n \left(\sum_r \overline{a_{\ell_1,r} a_{\ell_2,r}} a_{\ell_3,r} a_{\ell_4,r} \right) \left(\sum_k \overline{h_{\ell_1, k} h_{ \ell_2,k}} h_{ \ell_3,k} h_{ \ell_4,k}\right) = 0 $$ 
for all $\ell_1,\ell_2,\ell_3$ and $\ell_4$ such that $\{\ell_1,\ell_2\}\ne \{\ell_3,\ell_4\}$. 

Let us define the {\em L-graph} (denoted $L(A)$) of a flat matrix $A$ as follows. It is a simple graph with the same set of vertices as $K(A)$. Two vertices are adjacent if and only if the corresponding vectors are orthogonal. The previous identity leads to the following observation: 
\begin{prop} 
\label{grafy} 
The homogeneous system given by~(\ref{homogenous}) is a complete system of MUHs if and only if the graphs $L(A)$ and $L(H)$ together cover the complete graph. 
\end{prop} 

If matrices $A$ and $H$ satisfy the conditions of Proposition~\ref{grafy} and $A'$ and $H'$ are such matrices that $L(A)$ is a subgraph of $L(A')$ and the same holds for $L(H)$ and $L(H')$, then $A'$ and $H'$ also give rise to a complete system of MUHs via~(\ref{homogenous}). Hence, without loss of generality we may consider only matrices with maximal L-graphs. We will call them L-maximal flat or Hadamard matrices, respectively. What are they? We can say little on the subject at the moment, it is a topic for a future research.  

\begin{open} 
\label{maximal} 
Describe L-maximal flat and Hadamard matrices and the corresponding graphs. 
\end{open} 

An answer to this question would possibly allow a systematization of all complete homogeneous systems of MUBs. Anyway, it is already clear that L-maximal flat matrices cover L-maximal Hadamard matrices (because the latter is a special case of the former).

There is an important class of L-maximal Hadamard matrices. It is a very common example of Hadamard matrices and it is used in all known constructions of maximal families of MUBs. These are Fourier matrices which we will now introduce. 

\section{Fourier Matrices}  
Fourier matrix is the most popular type of Hadamard matrices. It is called so because it performs the Fourier transform of a finite Abelian group. Fourier transform is widely used in many areas of mathematics, physics and computer science. However, here we will be mostly interested in one simple property of Fourier matrices. Namely, the rows of the Fourier matrix of a group $G$ with the Schur product operation form a group isomorphic to the group $G$ (see later). 

Let us take an Abelian group $G = \mathbb{Z}_{d_1}\times \mathbb{Z}_{d_2}\times \cdots \times \mathbb{Z}_{d_m}$ of order $n=d_1d_2\cdots d_m$. By the structure theorem for finite Abelian groups, each finite Abelian group is isomorphic to a group of this form (see, e.g.,~\cite{hall}).    

Later on we will be also interested in the group $\tilde G = \mathbb{R}_{d_1}\times \mathbb{R}_{d_2}\times \cdots \times \mathbb{R}_{d_m}$, where $\mathbb{R}_a$ is the group of real numbers modulo $a$ with the addition operation. Note that $G_1\cong G_2$ does not imply $\tilde G_1\cong \tilde G_2$. Also, the group $G$ is a subgroup of $\tilde G$. In addition to that we will use notation $G^*$ for the set of non-zero elements of $G$, and $\tilde G^*$ for the set of elements of $\tilde G$ with at least one component being a non-zero integer.

The Fourier transform usually is defined via the dual group which is formed of all the characters of the group. A character of an Abelian group is its morphism to the multiplicative group of unit-modulus complex number. It is possible to establish an isomorphism from $G$ to $\hat{G}$ (the dual group) by 
\begin{equation} 
\label{harakter} 
\chi_a(b) = \exp(\sum_{j=1}^m \frac{2\pi i}{d_j} a_jb_j), 
\end{equation} 
where $a = (a_1,a_2,\dots,a_m)$ and $b=(b_1,b_2,\dots,b_m)$ are elements of $G$ and $\chi_a$ is the element of $\hat G$ corresponding to $a$. Note that the expression $\chi_a(b)$ is symmetric in $a$ and $b$. We will also extend the definition~(\ref{harakter}) to any $a$ and $b$ in $\tilde G$. The following lemma is a classical result.  

\begin{lem}   
\label{summa}   
Let $x$ be an element of $\tilde G$. Then    
$\sum\limits_{y\in G} \chi_y(x) = 0$   
if and only if $x\in \tilde G^*$.   
\end{lem} 

\pfstart   
Let us write $x=(x_1,x_2,\dots,x_m)$. We have:   
$$\sum_{y\in G} \chi_y(x) = \prod_{j=1}^m \sum_{k=0}^{d_j-1} \exp\left(\frac{2\pi i}{d_j} x_jk\right).$$   

Lemma follows from the fact that the roots of the equation  
$\sum\limits_{k=0}^{d_j-1}\omega^k=0$ in $\omega$ are exactly  
the roots of unity $\exp(\frac{2\pi i}{d_j} x_j)$, with $x_j$  
an integer, $0 < x_j < d_j$.  
\pfend   

\begin{col}
\label{FourierIsHadamard}
The matrix $F=(f_{i,j})$, indexed by the elements of $G$ and with $f_{i,j} = \chi_j(i)$, is an Hadamard matrix. 
\end{col} 

\pfstart 
Clearly, all elements of the matrix have absolute value 1. The inner product of the rows indexed by $a$ and $b$ with $a\ne b$ is 
$$\sum_{y\in G} \overline{\chi_y(a)}\chi_y(b) = \sum_{y\in G} \chi_y(b-a) = 0.$$ 
Hence, two distinct rows are orthogonal and the matrix $F$ is Hadamard. 
\pfend 

Matrix $F$ from the last corollary is called the {\em Fourier matrix} of the group $G$. As an example, if we take $G=\mathbb{Z}_n$, we obtain the matrix   
$$\begin{pmatrix}   
1 &1&1&\ldots&1\\   
1&\omega_n & \omega_n^2 & \ldots & \omega_n^{n-1}\\   
1&\omega_n^2 & \omega_n^4 & \ldots & \omega_n^{2n-2}\\   
\vdots&\vdots&\vdots&\ddots&\vdots\\   
1&\omega_n^{n-1} & \omega_n^{2n-2} & \ldots & \omega_n^{n^2-2n+1}   
\end{pmatrix}$$   
with $\omega_n = e^{2\pi i/n}$. An arbitrary Fourier matrix is equal to a tensor product of such matrices.

The Fourier matrix of the group G has some good properties. At first, it is symmetric. Next, let us denote by $R_{i}$ the row that corresponds to the element $i\in G$. It is easy to see that $R_i\circ R_j = R_{i+j}$. So, the set of rows (the set of columns) forms a group, with the Schur multiplication as an operation, that it is isomorphic to the original group $G$.  

{\em Remark. } Note that the statement of Corollary~\ref{FourierIsHadamard} holds in more general assumptions. Take any subset $X\subset \tilde G$ of size $|G|$ such that for any $a,b\in X$ with $a\ne b$ we have $a-b\in\tilde G^*$. Then the matrix $F=(f_{xj})$ ($x\in X$, $j\in G$), with $f_{xj} = \chi_j(x)$, is Hadamard. 

For example, if we take $G=\mathbb{Z}_3\times\mathbb{Z}_2$ and $X=\{(0,0), (0,1), (1,a), (1,1+a), (2,b), (2,1+b)\}$ where $0\le a,b\le 1$ are some reals, the we get a matrix equivalent to one in formula (4) of~\cite{six}. [Check!]

\section{Fourier Matrices in Homogeneous Systems} 
\label{FourierIn} 
Let $H$ be an Hadamard $n\times n$-matrix. Denote its rows by $\{R_i\}$, $i=0,1,\dots,n-1$. Rescaling of columns does not change the L-graph, so we may always assume that $R_0$ consists only of ones. 

For a fixed $i$ the set $\{R_i\circ R_j\mid j=0,\dots,n-1\}$ is an orthogonal basis of $\mathbb{C}^n$. Hence, any $R_a\circ R_b$ is not orthogonal to at least one of $\{R_i\circ R_j\mid j=0,\dots,n-1\}$. If $H$ is a Fourier matrix, then $R_a\circ R_b$ is not orthogonal to exactly one of $\{R_i\circ R_j\mid j=0,\dots,n-1\}$: the one with $i+j=a+b$.  

And conversely, if any $R_a\circ R_b$ is not orthogonal to exactly one of $\{R_0\circ R_j\mid j=0,\dots,n-1\}$ then $L(H)$ is isomorphic to the L-graph of a Fourier matrix. Indeed, let $G$ be the set of directions (equivalence classes of collinear vectors) defined by rows of $H$ with the Schur product operation. The set is finite, it is closed under the operation, $R_0$ is the identity element, the operation is commutative and associative and for any fixed $i$ the operation $R_j\mapsto R_i\circ R_j$ is a bijection. Hence, $G$ is a finite Abelian group, and $L(H)$ is isomorphic to the L-graph of the Fourier matrix of $G$. So, we have proved the following result 
\begin{thm} 
Fourier matrices are L-maximal Hadamard matrices. Moreover, their graphs have maximal possible number of edges. 
\end{thm} 

This result explains why Fourier matrices are so useful in the constructions of MUBs. It can be conjectured that Fourier matrices are the only L-maximal Hadamard matrices. 

Let $G$ be a graph. Recall~\cite{diestel} that the {\em independence number} $\alpha(G)$ is the greatest number of vertices that are pairwise disjoint, conversely, the {\em clique number} $\omega(G)$ is the greatest number of vertices that are all pairwise connected. The minimal number of colours that can be assigned to the vertices of the graph in such a way that any two adjacent vertices are coloured in different colours, is called the {\em chromatic number}  $\chi(G)$ of the graph. It is easy to show that $\chi(G)\ge\omega(G)$.  

It seems worth to mention some constructions that are similar to the notion of $L$-graphs (i.e., when the adjacency relation on the set of vectors is generated using the orthogonality relation). One known to us example is {\em Hadamard graph} defined in~\cite{ito}. The set of vertices of the Hadamard graph $S(n)$ of order $n$ is the set of all $\pm1$-component vectors of length n, and two vectors are adjacent iff they are orthogonal. The famous Hadamard conjecture is equivalent to the statement that $\omega(S(4n)) = 4n$ for any positive integer $n$. 

It is clear, that for any Hadamard matrix $H$ acting on $\mathbb{C}^n$ the clique number of $L(H)$ is equal to $n$. If $L(H)$ is a subgraph of the $L$-graph of a Fourier matrix, then $\chi(L(H))=n$ (the colour of a vertex is the corresponding element of the group). However, it is proved in~\cite{frankl} that there is an exponential gap between $4n$ and $\chi(S(4n))$. So, it is quite possible that for some Hadamard matrix $H$ we would have $\chi(L(B))>n$, and this matrix cannot be covered by a Fourier matrix.

Another nice property of Fourier matrices (noted to be ``striking'' in~\cite{six}) is that any vector $v$, unbiased with respect both to the standard basis and a Fourier matrix, can be collected into a whole unbiased basis. It is easy to explain if one notices that a Fourier matrix $F$ is symmetric and, hence, also its columns $\{R_i\}$ form a group with the Schur multiplication as the operation. The vector $v$ can be extended to a basis $\{R_a\circ v\mid a\in G\}$, and   
$$|\langle R_b| R_a\circ v\rangle| = |\langle R_{b-a}|v\rangle| = \frac1{\sqrt{n}}.$$   

As mentioned above, all known constructions of complete systems of MUBs are built from Fourier matrices. In the light of Proposition~\ref{grafy} it seems a good choice, since Fourier matrices are L-maximal Hadamard matrices. Moreover, the L-graph of a Fourier matrix covers a fraction of roughly $\frac{n-1}n$ edges of the complete graph, so it remains to find a flat matrix (a more general notion) to cover the remaining fraction of $\frac1n$ edges (a less number of edges). It seems that it should be easy, but it is not. 

\begin{prop}
Let $H$ be the Fourier matrix of a group $G$ and $A$ be a flat matrix with rows $\{R_i\}_{i\in G}$. The system defined by~(\ref{homogenous}) is a complete set of MUHs if and only if for any non-zero $\Delta\in G$ the matrix $D_\Delta$ with rows from 
$$\{R_{i+\Delta}\circ R_i^{(-1)}\mid i\in G\}$$ 
is an Hadamard matrix. (Here $R_i^{(-1)}$ stands for the element-wise inverse of $R_i$. It is the Schur (-1)-st power). 
\end{prop} 

\pfstart 
Suppose we have a complete set of MUHs. It follows from Proposition~\ref{grafy} that 
\begin{equation} 
\label{uslovie1} 
\forall g_1,g_2,g_3,g_4\in G: \left.\begin{array}{c}g_1+g_2=g_3+g_4\\\{g_1,g_2\}\ne \{g_3,g_4\}\end{array}\right\} \Longrightarrow R_{g_1}\circ R_{g_2} \perp R_{g_3}\circ R_{g_4}. 
\end{equation} 

Clearly, each element of $D_\Delta$ is of absolute value 1. Let us take $g_1\ne g_3$. Then 
$$\langle R_{g_1+\Delta}\circ R_{g_1}^{(-1)}|R_{g_3+\Delta}\circ R_{g_3}^{(-1)}\rangle =  
\langle R_{g_1+\Delta}\circ R_{g_3}|R_{g_3+\Delta}\circ R_{g_1}\rangle.$$ 
Moreover, $(g_1+\Delta)+g_3 = (g_3+\Delta)+g_1$, $g_3\ne g_1$ and $g_3\ne g_3+\Delta$. Using~(\ref{uslovie1}) with $g_2=g_3+\Delta$ and $g_4=g_1+\Delta$, we have $R_{g_1+\Delta}\circ R_{g_1}^{(-1)}\perp R_{{g_2}+\Delta}\circ R_{g_2}^{(-1)}$. 

The proof of the converse statement is similar.
\pfend 

Note that Hadamard matrices are quite rare, and here from one flat matrix one should extract $n-1$ Hadamards. It explains, to some extent, why it is not so easy to find a convenient matrix $A$. In practice, matrices $D_\Delta$ are chosen to be (up to some equivalence) equal to the same Fourier matrix. Now we give three possible kinds of restrictions on $D_\Delta$ and describe the corresponding constructions in terms of functions acting from one Abelian group into another.

Suppose matrix $H$ (as in~(\ref{homogenous})) is the Fourier matrix of the group $G = \mathbb{Z}_{d_1}\times \mathbb{Z}_{d_2}\times \cdots \times \mathbb{Z}_{d_m}$ and let $N = \mathbb{Z}_{d'_1}\times \mathbb{Z}_{d'_2}\times \cdots \times \mathbb{Z}_{d'_{m'}}$ be the group of the same size. Suppose all matrices $D_\Delta$ are equal (up to a permutation of rows) to the Fourier matrix $F$ of $N$ and each row of $A$ (that we want to construct) is a row of $F$. Define the function $f:G\to N$ as assigning to the index of a row of $A$ the index of the row of $F$ that stands in this place. It is easy to see that this construction satisfies the condition of Proposition~\ref{grafy} if and only $f$ satisfies
\begin{equation}   
\label{uslovie} 
\forall g_1,g_2,g_3,g_4\in G:   
\left.   
\begin{array}{r}   
g_1+g_2=g_3+g_4 \\   
f(g_1) + f(g_2) = f(g_3) + f(g_4) 
\end{array}\right\}\Longrightarrow \{g_1,g_2\}=\{g_3,g_4\}. 
\end{equation} 

This is not the most general case. If we allow $D_\Delta$ to be equal to the matrix $F$ with row permuted and each column multiplied by $\chi_x(a)$ where $a$ is the index of the column and $x$ is some element of $\tilde N$ (that depends on $\Delta$), then we can take the matrix $A=(a_{\ell r})$, ($\ell\in G, r\in N$) defined by $a_{\ell r} = \chi_r(f(\ell))$, where function $f:G\to\tilde N$ satisfies 
\begin{equation}   
\label{uslovieGeneral}   
\forall g_1,g_2,g_3,g_4\in G:
\left.   
\begin{array}{r}   
g_1+g_2=g_3+g_4 \\   
\{g_1,g_2\}\ne\{g_3,g_4\}   
\end{array}\right\}\Longrightarrow f(g_1) + f(g_2) - f(g_3) - f(g_4) \in N^*.   
\end{equation}   

Finally, from Lemma~\ref{summa} it follows that this approach gives a complete system of MUHs if and only if $f:G\to\tilde N$ satisfies
\begin{equation}   
\label{uslovieMostGeneral}   
\forall g_1,g_2,g_3,g_4\in G:
\left.   
\begin{array}{r}   
g_1+g_2=g_3+g_4 \\   
\{g_1,g_2\}\ne\{g_3,g_4\}   
\end{array}\right\}\Longrightarrow f(g_1) + f(g_2) - f(g_3) - f(g_4) \in \tilde N^*.   
\end{equation}   
However, in this case matrices $D_\Delta$ are not longer equivalent to a Fourier matrix, but rather to a matrix mentioned in the remark after Corollary~\ref{FourierIsHadamard}. 

Summing everything up, we have the following result: 
\begin{thm} 
\label{konstrukcija} 
Condition~(\ref{uslovieMostGeneral}) is more general than the one in~(\ref{uslovieGeneral}) that, in its turn, is more general than the one in~(\ref{uslovie}). Formula 
\begin{equation}   
\label{MUBformula}   
(v^{(r)}_k)_\ell = \frac{1}{\sqrt{n}}\chi_k(\ell)\chi_r(f(\ell)),   
\end{equation}   
(with $k,\ell\in G$ and $r\in N$) gives a complete system of MUHs if and only if the function $f:G\to\tilde N$ satisfies~(\ref{uslovieMostGeneral}). 
\end{thm} 

A similar result appeared in~\cite{roy}. We postpone a discussion of related topics to section~\ref{comb}. In the next section we show classical constructions of complete systems of MUBs in the light of Theorem~\ref{konstrukcija}. 

\section{Known Constructions}   
\label{known}   
Now we will give two known examples of complete sets of MUBs in the terms of the previous corollary.  

Construction essentially corresponding to the following one was first obtained for $GF(p)$ by Ivanovi\'c in~\cite{ivanovic} and in the general case by Fields and Wootters in~\cite{wootters}. 

\begin{lem}   
\label{ne4jot}  
If $n=p^k$ is a power of an odd prime, then the function  
$f(x) = x^2$ with $G=N$ being the additive group of $GF(n)$ (i.e. $\mathbb{Z}_p^k$) satisfies~(\ref{uslovie}).  
\end{lem}   

\pfstart   
Let us suppose $g_1+g_2=g_3+g_4$ and $g_1^2+g_2^2=g_3^2+g_4^2$. Then $g_1-g_3=g_4-g_2$ and $(g_1-g_3)(g_1+g_3) = (g_4-g_2)(g_4+g_2)$. If $g_1=g_3$, we are done. Otherwise, we can cancel $g_1-g_3$ out from the last equality and get $g_1+g_3 = g_4+g_2$. Together with the first equality it gives $2(g_2-g_3)=0$. Because 2 does not divide $n$, $g_2=g_3$ and we are done.   
\pfend   

If $n$ is even we have to be a bit more tricky. Let us remind, that the finite field $GF(2^k)$ consists of polynomials with degree smaller than $k$ and coefficients from $\{0,1\}$. All operations are performed modulo 2 and $h$, where $h$ is an irreducible polynomial of degree $k$. We will treat these polynomials as integer polynomials. The next lemma also leads to the construction first obtained by Fields and Wootters in~\cite{wootters}. 

\begin{lem}   
\label{4jot}   
Let $G$ be the additive group of $GF(2^k)$. Then the function $f:G\to\tilde G$ defined with    
$$f(x) = \frac{x^2}2 \bmod (2,h)$$   
satisfies~(\ref{uslovieGeneral}) with $N=G$.   
\end{lem}   

\pfstart   
Suppose $g_1+g_2\equiv g_3+g_4 \pmod{2,h}$. Then $(g_1+g_2)^2\equiv (g_3+g_4)^2\pmod{2,h}$. Hence, $g_1^2+g_2^2-g_3^2-g_4^2\equiv 0\pmod{2,h}$. This means that 
$$f(g_1)+f(g_2)-f(g_3)-f(g_4) = \frac{g_1^2+g_2^2-g_3^2-g_4^2}2\bmod (2,h)$$
is an integer polynomial. The only way it could not belong to $N^*$ is if it was equal to 0. Let us suppose it is equal to zero and prove that in this case $\{g_1,g_2\} = \{g_3,g_4\}$.

Let us define $s = (g_1+g_2) \bmod 2$. Then also $g_1^2+g_2^2 \equiv s^2 \pmod{2}$. Consider the following equation in $x$:   
$$\frac{g_1^2+g_2^2-x^2-(s-x)^2}{2}\equiv 0\pmod{h,2}.$$
Both $g_1$ and $g_2$ are its roots. The polynomial of $x$ can be rewritten as $\frac{g_1^2+g_2^2-s^2}{2}+sx-x^2$. One may notice that $\frac{g_1^2+g_2^2-s^2}{2}$ is an integer polynomial, so taking it modulo $h$ and 2 we obtain an equation of the second degree in $GF(2^k)$:   
$$x^2-sx-\frac{g_1^2+g_2^2-s^2}{2} = 0.$$   
If $g_1\ne g_2$, no other element except them can satisfy it. If $g_1=g_2$ then $s=0$ and this equation has only one root, because $x\mapsto x^2$ is a bijection in $GF(2^k)$ (a Frobenius map). Thus, $f$ satisfy~(\ref{uslovieGeneral}).
\pfend   

The last two lemmas combine into the following well-known result:   
\begin{thm}   
If $n$ is a prime power then there exists a complete set of MUBs in $\mathbb{C}^n$.   
\end{thm}   

\section{Related Combinatorial Structures} 
\label{comb} 
Observing formulas~(\ref{uslovie}),~(\ref{uslovieGeneral}) and~(\ref{uslovieMostGeneral}) one can conclude that they, especially~(\ref{uslovie}), are of a highly combinatorial nature. It turns out that they indeed have a strong link with some well-studied combinatorial structures. 

Suppose $G$ and $N$ are Abelian groups with $|G|\le |N|<\infty$.  
Functions $f: G\to N$, for which the equation $f(x+a)-f(x) = b$ has no more than 1 solution 
for all $a,b\in G$ not equal to zero simultaneously, are called {\em differentially 1-uniform}~\cite{nyberg}. If $N$ satisfies $|G|/|N| = m\in\mathbb{N}$ and function $f:G\to N$ is such that  $|\{x\in G\mid f(x+a)-f(x)=b\}|=m$ for any $b\in N$ and non-zero $a\in G$, the latter is called {\em perfect non-linear}~\cite{perfect}. These functions are used in cryptography to construct S-boxes that are not vulnerable to differential cryptanalysis. 

If $|G|=|N|$ as in~(\ref{uslovie}), these two notions coincide and function $f$ is sometimes called a {\em planar function}. This name is given because any planar function gives rise to an affine plane~\cite{planar}. For functions satisfying~(\ref{uslovieGeneral}) we will use name {\em fractional planar}. 

The following planar functions from $GF(p^k)$, with $p$ odd, to itself are known: 
\begin{itemize} 
	\item $f(x) = x^{p^\alpha+1}$, where $\alpha$ is a non-negative integer with $k/\gcd(k,\alpha)$ being odd. See~\cite{planar}. 
	\item $f(x) = x^{(3^\alpha+1)/2}$ only for $p=3$, $\alpha$ is odd, and $\gcd(k,\alpha)=1$. See~\cite{coulter}. 
	\item $f(x) = x^{10}-ux^6-u^2x^2$ only for $p=3$, $k$ is odd, and $u$ is a non-zero element of $GF(p^k)$. The special case of $u=-1$ was obtained in~\cite{coulter}, the general case is due to~\cite{ding}. 
\end{itemize} 
The construction with $f(x)=x^2$ from the previous section is from the first class. 

Let $K$ again be an Abelian group and $N$ be its subgroup. A subset $R\subset K$ is called a {\em relative $(m,n,r,\lambda)$-difference set} if $|K|=nm$, $|N|=n$, $|R|=r$ and 
$$|\{r_1,r_2\in R\mid r_1-r_2=b\}|=\left\{\begin{array}{rcl}nm&,&b=0; \\ 0&,&b\in N\setminus\{0\};\\ \lambda&,& b\in K\setminus N.\end{array}\right.$$ 
Relative difference set is a generalization of classical difference set and it was introduced in~\cite{relative}. If $r=m$ the difference set is called {\em semiregular}. A relative difference set is called {\em splitting} if $K=G\times N$, i.e. if $N$ has a complement in $K$. 

This notion is interesting to us because of the following easy observation (see, e.g., \cite{pott}). Let $G$ and $N$ be arbitrary finite groups and $f$ be a function from $G$ to $N$. The set $\{(x,f(x))\mid x\in G\}$ is a semiregular splitting $(|G|, |N|, |G|, |G|/|N|)$-difference set in $G\times N$ relative to $\{1\}\times N$ if and only if $f$ is perfect nonlinear. Thus, planar functions correspond to splitting relative $(n,n,n,1)$-difference sets. We extend this result a bit:

\begin{thm}
Let $K$ be an Abelian group of size $n^2$ having a subgroup $N = \mathbb{Z}_{d'_1}\times \mathbb{Z}_{d'_2}\times \cdots \times \mathbb{Z}_{d'_{m'}}$ of size $n$. The following two statements are equivalent:
\begin{itemize}
	\item[(a)] There exists a semiregular $(n, n, n, 1)$-difference set $R$ in $K$ relative to $N$.
	\item[(b)] There exists a fractional planar function $f: G\to \tilde N$ where $G\cong K/N$.
\end{itemize}
\end{thm} 

\pfstart 
Suppose we have a relative difference set. Fix any $G = \mathbb{Z}_{d_1}\times \mathbb{Z}_{d_2}\times \cdots \times \mathbb{Z}_{d_m}$ such that $G\cong N/K$. Let $g_1,\dots,g_m$ be elements of $K$ such that they, considered as elements of $G$, form the basis of the latter. Denote by $(s_{1i}, s_{2i}, \dots, s_{m'i})$ the element $d_ig_i\in N$, $i=1,\dots m$.

Define an Abelian group $K'$ as follows. Its elements are from the direct product $G\times N$ and the sum of two elements $(x_1,x_2,\dots,x_m; y_1,y_2,\dots,y_{m'})$ and $(z_1,z_2,\dots,z_m; t_1,t_2,\dots,t_{m'})$ is defined as $(a_1,a_2,\dots,a_m; b_1,b_2,\dots,b_{m'})$ where $a_i = x_i+z_i$ and 
\begin{equation}
\label{KPrim}
\begin{pmatrix}b_1\\b_2\\\vdots \\ b_{m'} \end{pmatrix} =  
\begin{pmatrix}y_1\\y_2\\\vdots \\ y_{m'} \end{pmatrix} + \begin{pmatrix}t_1\\t_2\\\vdots \\ t_{m'} \end{pmatrix} + 
\begin{pmatrix} 
s_{11} & s_{12} & \cdots & s_{1m}\\ 
s_{21} & s_{22} & \cdots & s_{2m}\\ 
\vdots & \vdots & \ddots & \vdots \\ 
s_{m'1} & s_{m'2} & \cdots & s_{m'm}\\ 
\end{pmatrix} 
\begin{pmatrix}[x_1+z_1\ge d_1]\\ [x_2+z_2\ge d_2] \\ \vdots \\ [x_m+z_m\ge d_m] \end{pmatrix}
\end{equation}
where $[x_i+z_i\ge d_i]$ is equal to 1 if the sum of $x_i$ and $z_i$, taken as integers, exceeds $d_i$ and is equal to 0 otherwise. It is not hard to check that $\varphi: K'\to K$, defined with $$(x_1,x_2,\dots,x_m; y) \mapsto y + \sum_{i=1}^m x_ig_i,$$ is an isomorphism. As usually, we identify elements of $G$ with the set $\{(x,0)\mid x\in G\}$ and $N$ with $\{(0;y)\mid y\in N\}$.

Denote by $S$ the $m'\times m$-matrix whose $(i,j)$-th element is equal to $s_{ij}/d_i$. Clearly, $\psi: K'\to \tilde N$ defined by $$\psi(x,y) = y+Sx$$ is a morphism. Since $R$ is a semiregular relative difference set, for any $x\in K/N$ we can find a unique element $r_x\in R$ with projection on $K/N$ equal to $x$. Define $f(x) = \psi(\varphi^{-1}(r_x))$. 

Let us prove that $f$ is fractional planar. At first, note that for any $x\in G$: $\varphi^{-1}(r_x) = (x,y)$ for some $y\in N$. Then suppose that $g_1,g_2,g_3,g_4\in G$ are such that $g_3-g_1 = g_2-g_4\ne 0$ and $g_1\ne g_4$. Denote $(x_1, y_1) = \varphi^{-1} (r_{g_3}-r_{g_1})$ and $(x_2,y_2) = \varphi^{-1}(r_{g_2}-r_{g_4})$. We have $x_1=x_2$ (because $g_3-g_1 = g_2-g_4$) and $y_1\ne y_2$ (because $r_{g_3}-r_{g_1}\ne r_{g_2}-r_{g_4}$). From the definition of $\psi$ we have $(f(g_3)-f(g_1))-(f(g_2)-f(g_4)) \in N^*$.

Suppose conversely that we have a fractional planar function $f:G\to\tilde N$ with the same expressions for $G$ and $N$. Define function $\{\cdot\}$ that takes a fractional part of every component of an element of $\tilde N$. Define also $\tilde f(x) =\{f(x)\}$. Then~(\ref{uslovieGeneral}) yields 
$$(a+b = c+d)\Longrightarrow (\tilde f(a)+\tilde f(b) - \tilde f(c) - \tilde f(d)\in N).$$ 

Since the condition on $f$ is invariant under adding a constant to the function, we may assume that $\tilde f(0)=0$. Then $\tilde f(a+b) = \{\tilde f(a)+\tilde f(b)\}$. Now it is easy to deduce that $\tilde f(x) = \{Sx\}$ where $S$ is defined in the same way as before for some integers $s_{ij}$. 

Define $K'$ as in~(\ref{KPrim}) and define $$R = \{(x;f(x)-Sx) \mid x\in G\}.$$
Similar reasoning as before shows that $R$ is semiregular difference set relative to $N$.
\pfend 

So, we have proved that if matrix $H$ in~(\ref{homogenous}) is a Fourier matrix, and all $D_\Delta$ are equivalent (in some sense) Fourier matrices, then the existence of a complete system of MUHs in $\mathbb{C}^n$ is equivalent to the existence of a relative $(n,n,n,1)$-difference set. In fact, a more general result~\cite{godsil} is known: the existence of a relative $(n,k,n,\lambda)$-difference set implies the existence of $k$ MUHs in $\mathbb{C}^n$. 

It is proved in~\cite{primepower} that a relative $(n,n,n,1)$-difference set exists only if $n$ is a prime power. Thus, using the approach with $f$ satisfying~(\ref{uslovieGeneral}) it is not possible to construct a complete system of MUBs for any new dimension. It is still not clear what can be said in the case of general $D_\Delta$ and, in particular, in the case of $f$ satisfying~(\ref{uslovieMostGeneral}).

\section{Conclusion} 
In this paper we have shown that MUBs stand close to sequences with low correlation, similar constructions and lower bounds can be used in both. In particular, one of the lower bounds (the Welch bounds) gives a nice characterisation of MUBs in terms of orthogonality of a certain collection of vectors. It could be interesting to try to use other constructions and bounds from one area in another.  

In particular, it is tempting to use criterion of Theorem~\ref{criterion} to other objects. One example could be SIC-POVMs, because it is proved in~\cite{are} that they also attain the Welch bound for $k=2$. However, it is not clear what other constraints should be added to a set of vectors to be a SIC-POVM (like being a union of bases in the case of MUBs). 

Our criterion seems to have no use for non-complete systems of MUBs. However, in~\cite{roy} it is proposed to use weighted 2-designs consisting of bases for quantum state estimation when no complete system of MUBs is known. Our criterion is suitable in this case as well. Let us give some details. 

The problem is to find such orthonormal bases $B_0,B_1,\dots,B_k$ of $\mathbb{C}^n$ and weights $w_0,w_1,\dots,w_k$ that are non-negative real numbers that the set $\{w_ix \mid x\in B_i, i=0,1,\dots,k\}$ attain the Welch bound for $k=2$. In particular, one of the main results of~\cite{roy} can be proved similarly to Theorem~\ref{konstrukcija}: 
\begin{thm} 
The existence of a differentially 1-uniform function $f$ from an Abelian group $G$ into an Abelian group $N$ with $|G|=n$ and $|N|=m$ implies the existence of a weighted 2-design in $\mathbb{C}^n$ formed from $m+1$ orthonormal bases. 
\end{thm} 

Another direction of future research is the investigation of L-maximal flat and Hadamard matrices in order to find new systems of MUBs or to prove that such systems can be reduced to other already studied cases. A question of non-homogeneous systems of MUBs also remains open. 

\section*{Acknowledgements} 
We would like to thank Jos\'e Ignacio Rosado for bringing our attention to some errors in the first version of the paper.

\end{document}